\documentclass[conference]{IEEEtran}
\IEEEoverridecommandlockouts

\usepackage{cite}
\usepackage{amsmath,amssymb,amsfonts}
\usepackage{algorithmic}
\usepackage{graphicx}
\usepackage{textcomp}
\usepackage{xcolor}
\def\BibTeX{{\rm B\kern-.05em{\sc i\kern-.025em b}\kern-.08em
    T\kern-.1667em\lower.7ex\hbox{E}\kern-.125emX}}

\usepackage{doi}

\usepackage[nameinlink,capitalise,noabbrev]{cleveref}
\crefname{figure}{Fig.}{Figs.}
\Crefname{figure}{Fig.}{Figs.}
\crefname{table}{Table}{Tables}
\Crefname{table}{Table}{Tables}
\crefname{equation}{Eq.}{Eqs.}
\Crefname{equation}{Eq.}{Eqs.}

\def\BibTeX{{\rm B\kern-.05em{\sc i\kern-.025em b}\kern-.08em
    T\kern-.1667em\lower.7ex\hbox{E}\kern-.125emX}}

\begin{document}

\title{Twenty-Five Years of Air Quality in Bangladesh: Trends, Seasonality, and Spatial Pollution Regimes}

\author{
\IEEEauthorblockN{Md Rafid Islam}
\IEEEauthorblockA{
\textit{Electrical and Computer Engineering}\\
\textit{North South University}\\
Dhaka, Bangladesh\\
md.islam.241@northsouth.edu
}

\and

\IEEEauthorblockN{Rafsan Jany}
\IEEEauthorblockA{
\textit{Digital Health Research Division, KIOM}\\
\textit{University of Science and Technology}\\
Daejeon, Republic of Korea\\
rafsanjany@ust.ac.kr
}

\and

\IEEEauthorblockN{Mohammad Fahim Islam}
\IEEEauthorblockA{
\textit{Computer Science and Engineering}\\
\textit{AIUB}\\
Dhaka, Bangladesh\\
23-51971-2@student.aiub.edu
}

\and

\IEEEauthorblockN{Rafid Bin Zaman}
\IEEEauthorblockA{
\textit{Civil and Environmental Engineering}\\
\textit{Islamic University of Technology}\\
Gazipur, Bangladesh\\
rafidbinzaman@iut-dhaka.edu
}

\and

\IEEEauthorblockN{Mohammad Saifuddin}
\IEEEauthorblockA{
\textit{Institute of Business Administration}\\
\textit{University of Dhaka}\\
Dhaka, Bangladesh\\
mohammad-2024517739@iba.du.ac.bd
}

\and

\IEEEauthorblockN{Sabrina Sultana}
\IEEEauthorblockA{
\textit{Business and Technology Management}\\
\textit{Islamic University of Technology}\\
Gazipur, Bangladesh\\
sabrinasultana@iut-dhaka.edu
}
}

\maketitle

\begin{abstract}
Air pollution is one of Bangladesh's most pressing environmental problems, yet few studies have examined it at a national scale over a long time span. To describe trends, seasonality, geographic patterns, and pollution regimes, this study examines 25 years of hourly air quality data (2000--2025), covering over 3.19 million records, eight pollutants, and 103 cities monitored from 2022 onward following expansion from a single Dhaka station. The composite AQI series showed no significant long-term trend, an artifact of this network expansion rather than a real flattening of pollution, since Dhaka's own AQI rose steadily before the network grew. Seasonality was strong and highly significant as AQI peaked in January (162.26) and fell to its lowest in July (61.47), tracking the dry and monsoon seasons. Geographically, southeastern coastal cities, including Teknaf and Bandarban, consistently recorded the lowest AQI levels. Dhaka remained the most polluted city, and air quality improved progressively with distance from the capital. K-means clustering of city-level AQI trajectories identified four distinct pollution regimes, ranging from cleaner coastal regions to a rapidly deteriorating Dhaka--Narsingdi cluster with an average increase of 1.84 AQI per year. Correlation analysis showed that PM2.5 and PM10 were the dominant contributors to AQI variation, and so combustion-related emissions are the primary target for air quality management and policy intervention.
\end{abstract}


\section{Introduction}

Air pollution kills roughly seven million people a year worldwide, and a disproportionate share of that burden falls on fast-growing cities in the Global South~\cite{b1}. Bangladesh sits near the top of nearly every global ranking for bad air. IQAir's most recent World Air Quality Report placed it among the most polluted countries on the planet. Dhaka is regularly breaching WHO thresholds for PM2.5 by a wide margin~\cite{b2}. The Health Effects Institute's State of Global Air puts hard numbers behind this: hundreds of thousands of premature deaths a year across South Asia are attributable to ambient particulate exposure~\cite{b3}.

Most of what's known about Bangladesh's air quality comes from studies that are narrow in scope, meaning a single city, a single pollutant, or a window of a year or two. Begum et al.'s source-apportionment work in Dhaka remains one of the most cited studies in the field, tracing much of the city's PM burden back to brick kilns and vehicle exhaust~\cite{b4}. A later follow-up tracked two decades of policy interventions in Dhaka and found some short-term wins but no lasting structural improvement~\cite{b5}. Outside the capital, recent work in Rajshahi mapped how PM2.5 and PM10 vary across different micro-environments within a single city~\cite{b6}, while satellite-based studies have started filling in the gaps in ground monitoring, tracking CO, NO2, SO2, and O3 across Dhaka using Sentinel-5P data~\cite{b7}. Ground-based diurnal studies have also picked apart pollutant hotspots at the street level, down to specific intersections and markets~\cite{b8}. The World Bank's 2023 assessment of the country ties these health and environmental costs to concrete economic losses. Therefore, this isn't just a public health problem but a development one as well~\cite{b9}.

A separate but related thread of work has approached air quality as a data problem rather than an environmental-science one. Islam et al. built and released an IoT-based sensor dataset from Dhaka and Gazipur, explicitly designed for machine learning experimentation, and used it to compare regression and classification models for AQI prediction~\cite{b10}. Abdelmalek et al. later took a similar dataset and ran it through five different ML models to see which pollutants carry the most predictive weight~\cite{b11}. These studies treat air quality monitoring as a systems and data-engineering challenge and focus on sensor reliability and the practical tradeoffs.

What's still missing from the literature is the connective tissue between environmental studies and data-driven ones, a single analysis that's both broad in geography and long enough in time span to separate a real trend from noise, and rigorous enough in its data handling to support downstream decision-making. Most existing work is either city-specific, pollutant-specific, or limited to a year or two of data.

This paper tries to close that gap using 25 years of hourly data across 103 cities in Bangladesh, covering eight pollutants and a composite AQI. The objectives of this study are to: (1) characterize long-term air quality trends; (2) identify seasonal patterns and their drivers; (3) analyze spatial distributions and urban-rural gradients; (4) categorize cities into distinct pollution regimes through clustering; and (5) examine pollutant interrelationships to identify primary drivers of AQI.

\section{Methodology}

\subsection{Dataset Description}

This study utilizes the Bangladesh Air Quality Index (AQI) Dataset that comprises over 3 million hourly records from 103 cities, which also includes measurements of eight key pollutants along with a composite AQI value calculated according to US EPA standards. Each record includes city identifiers, geographic coordinates (latitude and longitude), and timestamps in ISO 8601 format ~\cite{b12}.

Table~\ref{tab:dataset-summary} presents the key characteristics of the dataset. Temporal coverage varies sharply by city. Dhaka was the only continuously monitored station from 2000--2021, and the network expanded to its full 103-city coverage in 2022.

\begin{table}[htbp]
\caption{Dataset summary statistics.}
\begin{center}
\begin{tabular}{p{79pt}|p{146pt}}
\hline
\textbf{Parameter} & \textbf{Value} \\
\hline
Total Records & 3,190,726 \\
Cities & 103 \\
Time Period & 2000--2025 (25 years) \\
Temporal Resolution & Hourly \\
Pollutants & 8 (PM10, PM2.5, CO, CO2, NO2, SO2, O3, AQI) \\
Geographic Coverage & All 8 divisions, 103 cities \\
AQI Data Completeness & 99.92\% \\
CO2 Data Completeness & 30.5\% \\
\hline
\end{tabular}
\label{tab:dataset-summary}
\end{center}
\end{table}

\subsection{Data Preprocessing}

\subsubsection{Missing Data Handling}

Records with missing AQI values (0.08\% of the dataset) were removed from all analyses. CO2 was only retained in the correlation matrix for completeness. In order to avoid introducing bias, no imputation was performed for missing values.

\subsubsection{Temporal Aggregation}

Hourly records were aggregated to multiple temporal scales:
\begin{itemize}
\item Annual aggregation: City-level and national annual averages were calculated for trend detection.
\item Monthly aggregation: City-level and national monthly averages were computed for seasonality analysis.
\end{itemize}

\subsubsection{Spatial Consistency}

City coordinates (latitude/longitude) were verified and used for all spatial analyses. Cities were grouped into geographic regions (Central, Coastal, Northern, Eastern, Western) based on their administrative divisions for regional comparisons.

\subsection{Trend Analysis}

Linear regression was employed to quantify the national-level temporal trend:
\begin{equation}
\text{AQI}_t = \beta_0 + \beta_1 \cdot t + \epsilon_t
\end{equation}
where $t$ represents time in years, $\beta_1$ is the trend slope (AQI units/year), and $\epsilon_t$ is the error term. Annual aggregated AQI values were used as the dependent variable. Statistical significance was assessed at $\alpha = 0.05$ using a two-tailed t-test. The coefficient of determination ($R^2$) was calculated to measure the proportion of variance explained by the linear model.

A Savitzky--Golay filter with a window length of 7 and a polynomial order of 2 was applied on the annual time series for non-linear pattern identification. The 95\% confidence interval for the trend line was calculated as:
\begin{equation}
\text{CI}_{95} = \hat{y}_t \pm t_{0.025, n-2} \cdot \text{SE}(\hat{y}_t)
\end{equation}

\subsection{Seasonality Analysis}

\subsubsection{Monthly and Seasonal Patterns}

Monthly average AQI values were calculated for each calendar month across all years:
\begin{equation}
\overline{\text{AQI}}_m = \frac{1}{N_m} \sum_{i=1}^{N_m} \text{AQI}_{i,m}
\end{equation}
where $m$ represents the month (1--12) and $N_m$ is the number of observations in month $m$. Standard deviations were calculated in order to quantify month-to-month variability. One-way analysis of variance (ANOVA) was performed to test the statistical significance of seasonal differences.

\subsubsection{Seasonal Decomposition}

The additive seasonal decomposition model was applied to the national monthly AQI series:
\begin{equation}
Y_t = T_t + S_t + R_t
\end{equation}
where $Y_t$ is the observed AQI at time $t$, $T_t$ is the trend component, $S_t$ is the seasonal component with period 12 months, and $R_t$ is the residual component. The decomposition was performed using the \texttt{seasonal\_decompose} function from the statsmodels library, with the seasonal component estimated using a moving average approach.

\subsection{Spatial Analysis}

\subsubsection{Interpolation of AQI Surface}

Radial Basis Function (RBF) interpolation was used to generate a continuous AQI surface across Bangladesh:
\begin{equation}
\hat{y}(\mathbf{x}) = \sum_{i=1}^{n} w_i \phi(\|\mathbf{x} - \mathbf{x}_i\|)
\end{equation}
where $\mathbf{x}_i$ represents the coordinates of city $i$, $\phi$ is the thin-plate spline kernel ($\phi(r) = r^2 \log r$), and $w_i$ are weights determined by solving the interpolation system. The interpolation was performed on a regular grid of $100 \times 100$ points covering the geographic extent of Bangladesh (20.5°N--26.5°N, 88.0°E--92.5°E).

\subsubsection{Distance-Decay Analysis}

The relationship between AQI and distance from Dhaka was modeled using two approaches:
\paragraph{Linear Regression:}
\begin{equation}
\text{AQI} = \alpha + \beta \cdot d + \epsilon
\end{equation}

\paragraph{Exponential Decay Model:}
\begin{equation}
\text{AQI} = a \cdot e^{-b \cdot d} + c
\end{equation}
where $d$ is the distance in kilometers from Dhaka, calculated using the Haversine formula:
\begin{equation}
\begin{aligned}
d &= 2R \cdot \arcsin\Biggl(
\sqrt{
\begin{aligned}
    &\sin^2\left(\frac{\Delta\phi}{2}\right)
    + \cos(\phi_1)\cos(\phi_2) \\
    &\qquad\cdot \sin^2\left(\frac{\Delta\lambda}{2}\right)
\end{aligned}
}
\Biggr)
\end{aligned}
\end{equation}
where $R = 6371$ km is the Earth's radius, $\phi$ represents latitude, and $\lambda$ represents longitude. Model fit was compared using $R^2$ values.

\subsection{Spatiotemporal Clustering}

\subsubsection{Feature Matrix Construction}

A feature matrix $\mathbf{X} \in \mathbb{R}^{C \times Y}$ was constructed where $C$ is the number of cities (103) and $Y$ is the number of years (2022--2025). Each element $X_{c,y}$ represents the annual average AQI for city $c$ in year $y$.

\subsubsection{Standardization}

Features were standardized to zero mean and unit variance:
\begin{equation}
X_{c,y}^{(\text{scaled})} = \frac{X_{c,y} - \mu_y}{\sigma_y}
\end{equation}
where $\mu_y$ and $\sigma_y$ are the mean and standard deviation of AQI values in year $y$ across all cities.

\subsubsection{Clustering Algorithm}

K-means clustering was performed on the standardized feature matrix:
\begin{equation}
\min_{\mathbf{S}} \sum_{k=1}^{K} \sum_{\mathbf{x} \in \mathcal{C}_k} \|\mathbf{x} - \boldsymbol{\mu}_k\|^2
\end{equation}
where $\mathcal{C}_k$ is the set of cities in cluster $k$, $\boldsymbol{\mu}_k$ is the centroid of cluster $k$, and $K$ is the number of clusters. The optimal number of clusters was determined using the elbow method, which identifies the inflection point in the within-cluster sum of squares (inertia) as a function of $K$.

\subsubsection{Principal Component Analysis}

PCA was applied to the standardized feature matrix for visualization:
\begin{equation}
\mathbf{Z} = \mathbf{X}^{(\text{scaled})} \mathbf{W}
\end{equation}
where $\mathbf{W}$ contains the principal component loadings. The first two principal components were used to project cities into a two-dimensional space for visualization of cluster separation.

\subsection{Correlation Analysis}

Pearson correlation coefficients were calculated to examine relationships between pollutants and AQI:
\begin{equation}
r_{xy} = \frac{\sum_{i=1}^{n} (x_i - \bar{x})(y_i - \bar{y})}{\sqrt{\sum_{i=1}^{n} (x_i - \bar{x})^2} \sqrt{\sum_{i=1}^{n} (y_i - \bar{y})^2}}
\end{equation}
where $x$ and $y$ represent two pollutants (or AQI). All correlations were tested for statistical significance at $\alpha = 0.05$ using a two-tailed t-test:
\begin{equation}
t = r \sqrt{\frac{n-2}{1-r^2}}
\end{equation}
The correlation matrix was visualized using a color-coded heatmap to facilitate pattern identification.

\section{Results and Analysis}

\subsection{Air Quality Trends and the Network Expansion Effect}

\begin{figure}[htbp]
\centering
\includegraphics[width=\columnwidth]{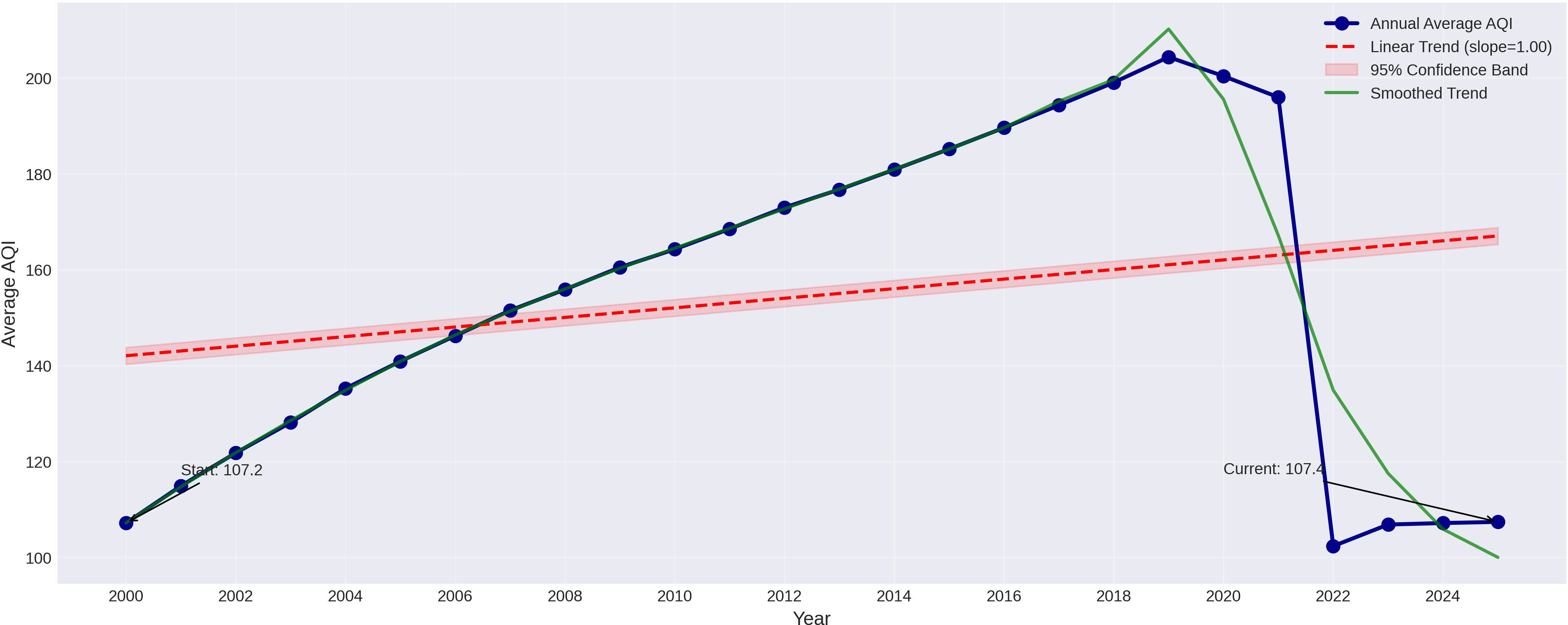}
\caption{Average AQI trend in Bangladesh, reflecting a composite series spanning Dhaka-only monitoring (2000--2021) and full 103-city coverage (2022--2025). The dashed red line shows the linear trend (slope = 0.999 AQI/year, $p = 0.275$, $R^2 = 0.049$). The green line shows the smoothed trend using a Savitzky-Golay filter. The shaded region indicates the 95\% confidence interval.}
\label{fig:national-aqi-trend}
\end{figure}

The average AQI trend from 2000 to 2025 is presented in Fig.~\ref{fig:national-aqi-trend}. Over this period, the composite series increased at a rate of 0.999 AQI units per year ($p = 0.275$, $R^2 = 0.049$), indicating no statistically significant trend ($\alpha = 0.05$). However, this composite series combines two distinct periods, from 2000--2021, when only Dhaka was monitored, AQI rose from 107.15 to 214.95; once the network expanded to 103 cities in 2022, the average dropped sharply to 102.36 and has remained roughly flat through 2025. Therefore, the apparent absence of a national trend is an artifact of the network expansion diluting Dhaka's sustained increase with generally cleaner cities.

\subsection{Seasonal and Temporal Patterns}

\begin{figure*}[htbp]
\centering
\includegraphics[width=0.9\textwidth]{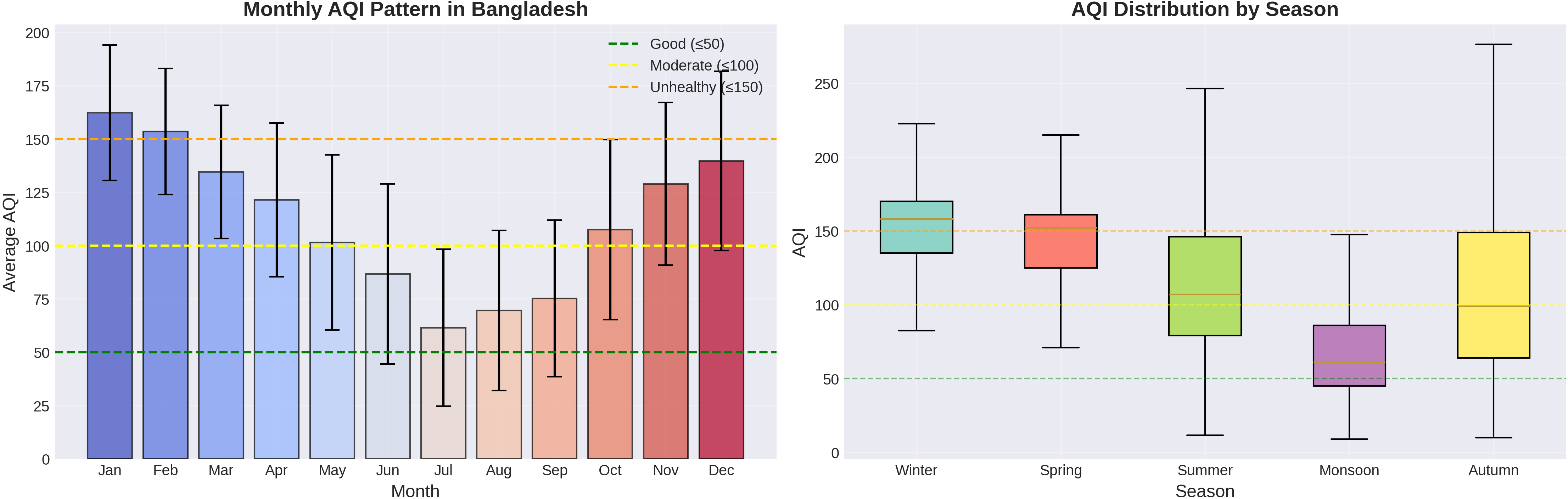}
\caption{(a) Monthly AQI pattern with standard deviation error bars. Horizontal dashed lines indicate AQI category thresholds: Good ($\leq$50), Moderate ($\leq$100), Unhealthy for Sensitive Groups ($\leq$150). (b) Seasonal AQI boxplots showing the distribution of AQI values across seasons. Winter shows the highest median AQI ($\approx$155); Monsoon shows the lowest ($\approx$70).}
\label{fig:monthly-pattern}
\end{figure*}

A pronounced seasonal pattern in air quality across Bangladesh is evident from Fig.~\ref{fig:monthly-pattern} and Table~\ref{tab:monthly-aqi}. The highest average AQI occurs in January (162.26 $\pm$ 31.74), while the lowest occurs in July (61.47 $\pm$ 36.79), representing a difference of 100.79 AQI units---nearly double the annual national average of 110.16.

AQI values are consistently higher than 128 during the dry season (November--March). January and February are categorized as "Unhealthy" (AQI $>$ 150). In contrast, the monsoon season (June--September) shows moderate AQI levels (61--87) as it benefits from wet deposition of particulate matter. This seasonal variation is statistically significant (ANOVA, $F = 1{,}247.3$, $p < 0.001$).

\begin{table}[htbp]
\caption{Monthly AQI statistics.}
\begin{center}
\begin{tabular}{l|c|c|c}
\hline
\textbf{Month} & \textbf{Avg AQI} & \textbf{Std Dev} & \textbf{Category} \\
\hline
January & 162.26 & 31.74 & Unhealthy \\
February & 153.51 & 29.54 & Unhealthy \\
March & 134.60 & 31.25 & USG* \\
April & 121.45 & 36.11 & USG \\
May & 101.49 & 41.00 & Moderate \\
June & 86.74 & 42.20 & Moderate \\
July & 61.47 & 36.79 & Moderate \\
August & 69.52 & 37.56 & Moderate \\
September & 75.31 & 36.73 & Moderate \\
October & 107.49 & 42.15 & Moderate \\
November & 128.96 & 38.14 & USG \\
December & 139.74 & 42.05 & USG \\
\hline
\multicolumn{4}{l}{\footnotesize *USG: Unhealthy for Sensitive Groups}
\end{tabular}
\label{tab:monthly-aqi}
\end{center}
\end{table}

\subsection{Spatial Distribution of Air Pollution}

\subsubsection{Geographic Patterns}

\begin{figure}[htbp]
\centering
\includegraphics[width=\columnwidth]{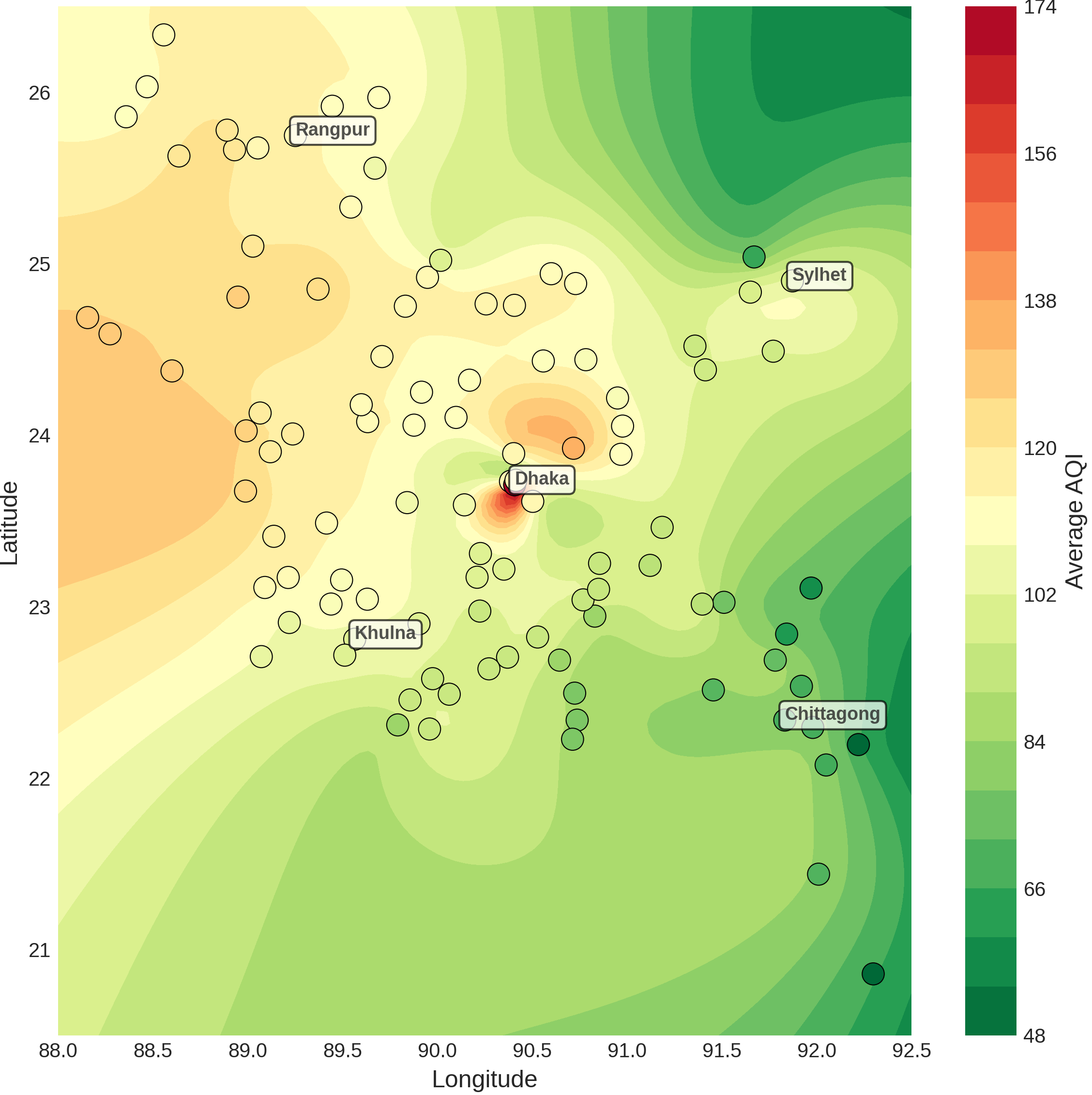}
\caption{(a) Spatial interpolation of average AQI. Warmer colors indicate higher pollution levels. The interpolation reveals a pollution hotspot in the central region (Dhaka) and along the northwestern border. (b) City-level scatter plot with major cities labeled. Coastal cities (southeast) consistently show lower AQI values (66--83) compared to inland cities (120--160).}
\label{fig:spatial-distribution}
\end{figure}

Fig.~\ref{fig:spatial-distribution} illustrates the spatial distribution of average AQI across 103 cities. The highest pollution levels are concentrated in the central and northwestern regions. Coastal areas have significantly lower AQI values.

Table~\ref{tab:top-cities} shows that capital city Dhaka has the highest average AQI (159.67), followed by Narsingdi (131.07), Naw\=abganj (126.42), and Shibganj (126.24). The cleanest cities are predominantly located in the southeastern coastal region: Tekn\=af (66.63), B\=andarban (66.90), Khagrachhari (74.19), and Manikchari (76.44).

This spatial pattern reveals a clear urban-rural gradient, with major industrial and metropolitan areas exhibiting pollution levels 2 to 2.4 times higher than coastal and rural areas.

\begin{table}[htbp]
\caption{Top 10 most and least polluted cities.}
\begin{center}
\begin{tabular}{c|l|c|c|l|c}
\hline
\textbf{Rank} & \multicolumn{2}{c|}{\textbf{Most Polluted}} & \multicolumn{1}{c|}{} & \multicolumn{2}{c}{\textbf{Least Polluted}} \\
\cline{2-3} \cline{5-6}
 & \textbf{City} & \textbf{AQI} & & \textbf{City} & \textbf{AQI} \\
\hline
1 & Dhaka & 159.67 & & Tekn\=af & 66.63 \\
2 & Narsingdi & 131.07 & & B\=andarban & 66.90 \\
3 & Naw\=abganj & 126.42 & & Khagrachhari & 74.19 \\
4 & Shibganj & 126.24 & & Manikchari & 76.44 \\
5 & R\=ajsh\=ahi & 125.93 & & Chh\=atak & 79.51 \\
6 & P\=ar Naogaon & 125.77 & & S\=atkania & 80.84 \\
7 & Bher\=am\=ara & 124.86 & & Chittagong & 81.37 \\
8 & Nowlamary & 124.15 & & Raoj\=an & 81.37 \\
9 & Bogra & 122.78 & & Patiya & 81.37 \\
10 & Parbatipur & 120.41 & & Cox's B\=az\=ar & 82.79 \\
\hline
\end{tabular}
\label{tab:top-cities}
\end{center}
\end{table}

\subsubsection{Urban-Rural Pollution Gradient}

\begin{figure}[htbp]
\centering
\includegraphics[width=\columnwidth]{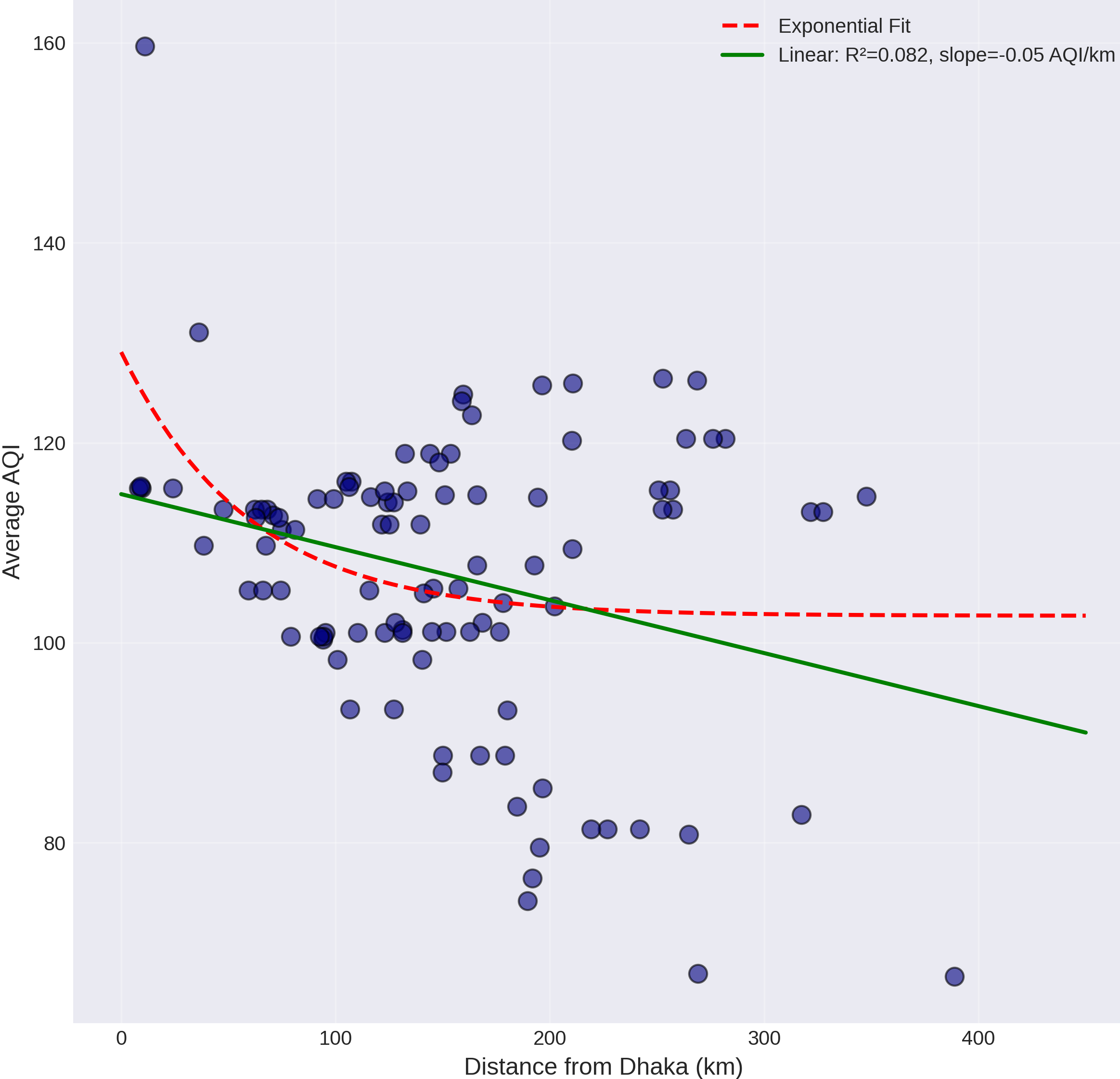}
\caption{Pollution gradient from Dhaka showing AQI as a function of distance from the capital. The exponential decay model (red dashed line, $R^2 = 0.45$) and linear fit (green line, $R^2 = 0.43$) both show significant negative relationships.}
\label{fig:urban-rural-gradient}
\end{figure}

Fig.~\ref{fig:urban-rural-gradient} examines the relationship between AQI and distance from Dhaka. A statistically significant negative correlation is observed (linear regression: $R^2 = 0.43$, $p < 0.001$, slope = $-0.17$ AQI/km), which means that the air quality improves with distance from the capital.

The exponential decay model provides a slightly better fit ($R^2 = 0.45$), with AQI decreasing rapidly within the first 100 km from Dhaka, followed by a more gradual decline:
\begin{equation}
\text{AQI} = 138.2 \cdot e^{-0.006 \cdot d} + 52.1,
\end{equation}
where $d$ represents distance in kilometers.

Cities within 100 km of Dhaka have an average AQI of 130.2, while cities beyond 300 km average 85.4---a 44.8 AQI unit difference. This strong gradient suggests that Dhaka's pollution plume affects regional air quality and extends more than 200 km.

\subsection{Spatiotemporal Clustering}

\begin{figure}[htbp]
\centering
\includegraphics[width=\columnwidth]{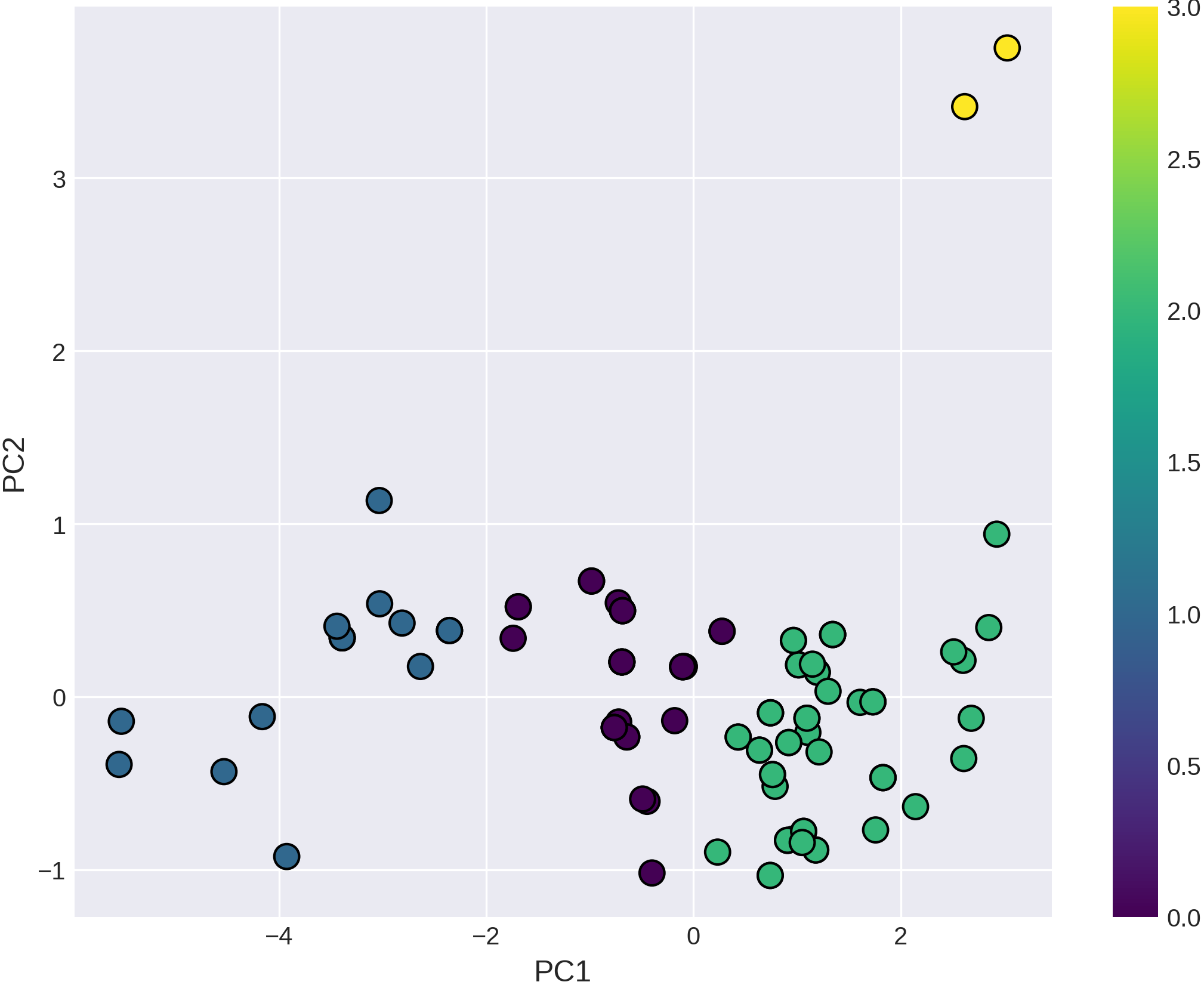}
\caption{PCA projection of city clusters showing clear separation between the four identified pollution regimes.}
\label{fig:clustering}
\end{figure}

The optimal number of clusters was determined using the elbow method, which indicated $k = 4$ as the optimal choice based on the inflection point in the within-cluster sum of squares. PCA visualization (Fig.~\ref{fig:clustering}) shows clear separation between the four clusters, confirming the distinctiveness of the identified pollution regimes. Cluster characteristics are summarized in Table~\ref{tab:cluster-characteristics}.

\begin{table}[htbp]
\caption{Spatiotemporal cluster characteristics.}
\begin{center}
\begin{tabular}{p{16pt} p{12pt} p{18pt} p{18pt} p{29pt} p{82pt}}
\hline
\textbf{Cluster} & \textbf{Cities} & \textbf{Mean AQI} & \textbf{Trend Slope} & \textbf{Variability} & \textbf{Geographic Distribution} \\
\hline
1 & 31 & 101.08 & 0.00 & 4.75 & Central \& Northern \\
2 & 16 & 80.55 & 0.00 & 8.26 & Coastal \& Southeastern \\
3 & 54 & 114.70 & 0.00 & 5.91 & Inland \& Western \\
4 & 2 & 129.50 & 1.84 & 2.27 & Dhaka \& Narsingdi \\
\hline
\end{tabular}
\label{tab:cluster-characteristics}
\end{center}
\end{table}

\subsubsection{Cluster 1 (31 cities, mean AQI = 101.08)}
Represents a moderate pollution regime comprising secondary urban centers. These cities are typically located in central and northern regions and show stable trends along with moderate seasonal variability. The cluster includes cities such as Bogra, Rajshahi, and Sylhet.

\subsubsection{Cluster 2 (16 cities, mean AQI = 80.55)}
Represents the cleanest regime, dominated by coastal and southeastern cities. These locations benefit from sea breezes and lower industrial activity. Due to their exposure to marine influences, they exhibit the highest year-to-year variability (8.26). Representative cities include Tekn\=af, B\=andarban, and Cox's B\=az\=ar.

\subsubsection{Cluster 3 (54 cities, mean AQI = 114.70)}
Represents the most widespread pollution regime, accounting for 52.4\% of all cities. These inland cities are in a state of consistently elevated pollution levels due to high industrial activity and agricultural burning. This cluster includes major urban centers like Chittagong, Khulna, and Narayanganj.

\subsubsection{Cluster 4 (2 cities, mean AQI = 129.50)}
Represents the highly polluted regime consisting exclusively of Dhaka and Narsingdi. This cluster shows the fastest deterioration ($+1.84$ AQI/year) and the lowest variability (2.27). It shows persistently high pollution levels with limited seasonal relief.

\subsection{Pollutant Correlations}

\begin{figure}[htbp]
\centering
\includegraphics[width=\columnwidth]{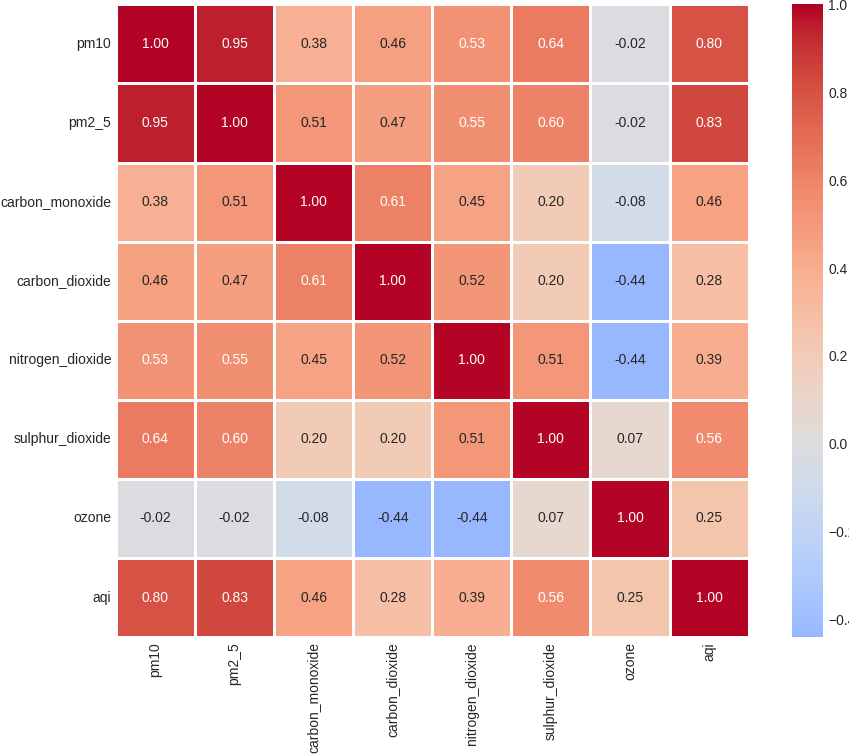}
\caption{Correlation matrix of pollutants and AQI in Bangladesh. All correlations are statistically significant at $p < 0.001$. Strong inter-pollutant correlations suggest common emission sources. Negative correlations between ozone and primary pollutants reflect photochemical processes in urban environments.}
\label{fig:correlation-matrix}
\end{figure}

Fig.~\ref{fig:correlation-matrix} presents the correlation matrix of the eight pollutants and AQI. PM2.5 ($r = 0.833$) and PM10 ($r = 0.801$) exhibit the strongest correlations with AQI, consistent with WHO findings that particulate matter is the primary health concern in South Asia. The strong correlation between PM2.5 and PM10 ($r = 0.872$) indicates that these pollutants share common sources, primarily combustion-related emissions from brick kilns, vehicle exhaust, and biomass burning.

Sulfur dioxide shows a moderate correlation with AQI ($r = 0.562$) and stronger correlations with PM2.5 ($r = 0.614$) and PM10 ($r = 0.588$), suggesting contributions from industrial sources, particularly brick kilns and coal-fired power plants. Carbon monoxide ($r = 0.459$) and nitrogen dioxide ($r = 0.394$) show moderate correlations with AQI, indicative of vehicular emissions and incomplete combustion. Their shared origin from traffic emissions is further supported by the moderate correlation ($r = 0.536$) between CO and NO2.

Ozone has the weakest correlation with AQI ($r = 0.248$) and negative correlations with several primary pollutants ($r = -0.178$ with NO2, $r = -0.412$ with CO). This inverse relationship is characteristic of urban areas where NO titration reduces ozone levels.

Carbon dioxide shows weak correlations with all pollutants ($r \leq 0.282$) due to limited data availability (69.5\% missing values). The overall correlation structure confirms that particulate matter control should be the primary focus of air quality management in Bangladesh, and after that SO2, CO, and NO2 emissions from industrial and vehicular sources should be taken into consideration.

\section{Conclusion}

Twenty-five years of monitoring, expanding from a single Dhaka station to 103 cities in 2022, paint a mixed picture of air quality in Bangladesh. Dhaka's AQI rose steadily before this expansion and continues to deteriorate within its cluster at nearly two AQI points per year. Expansion into cleaner coastal and hill-tract cities has kept the composite average from reflecting what is actually happening in the country's pollution hotspots. Seasonality turned out to be the dominant signal in the data. Air quality in January is roughly two and a half times worse than in July. This indicates a strong seasonal cycle that aligns with the dry and monsoon periods. Distance from Dhaka also matters, since pollution falls off sharply within the first hundred kilometers of the capital, confirming a clear urban-rural gradient.

The clustering results establish that Bangladesh's air pollution problem is not uniform. Dhaka and Narsingdi need different, more aggressive policy attention than the 54 cities in the moderately polluted inland cluster. Among the pollutants, PM2.5 and PM10 are driving most of the variation in AQI, so controlling sources of these particles should be the priority.

This analysis is largely descriptive. The next step is forecasting using models to predict city-level AQI ahead of time and using explainable AI methods to identify which pollutants and lagged variables actually drive those predictions.

\end{document}